\documentclass[manuscript,screen]{acmart}\usepackage[]{graphicx}\usepackage[]{xcolor}
\makeatletter
\def\maxwidth{ %
  \ifdim\Gin@nat@width>\linewidth
    \linewidth
  \else
    \Gin@nat@width
  \fi
}
\makeatother

\definecolor{fgcolor}{rgb}{0.345, 0.345, 0.345}

\usepackage{framed}
\makeatletter
 {\par\unskip\endMakeFramed%
 \at@end@of@kframe}
\makeatother

\definecolor{shadecolor}{rgb}{.97, .97, .97}
\definecolor{messagecolor}{rgb}{0, 0, 0}
\definecolor{warningcolor}{rgb}{1, 0, 1}
\definecolor{errorcolor}{rgb}{1, 0, 0}

\usepackage{alltt}

\usepackage{multirow}
\usepackage{censor}
\usepackage{array}

\AtBeginDocument{%
  \providecommand\BibTeX{{%
    \normalfont B\kern-0.5em{\scshape i\kern-0.25em b}\kern-0.8em\TeX}}}

\setcopyright{acmlicensed}
\copyrightyear{2024}
\acmYear{2024}
\acmDOI{XXXXXXX.XXXXXXX}

\acmConference[Conference acronym 'XX]{Conference Title}{Month DD--DD,
  YYYY}{City, State}

\acmBooktitle{Conference Acronym 'YY: Full Conference Name,
 Month DD--DD, YYYY, City, State} 
\acmISBN{978-1-4503-XXXX-X/18/06}

\title{Measuring the Mental Health of Content Reviewers, a Systematic Review}

\author{Alexandra Gonzalez}
\authornote{Both authors contributed equally to this research.}
\email{ag2425@cornell.edu}

\author{J. Nathan Matias}
\authornotemark[1]
\email{nathan.matias@cornell.edu}
\orcid{0000-0001-8910-0208}

\affiliation{%
  \institution{Citizens \& Technology Lab, Cornell University}
  \city{Ithaca}
  \state{New York}
  \country{USA}
  \postcode{43017-6221}
}
\IfFileExists{upquote.sty}{\usepackage{upquote}}{}
\begin{document}

\renewcommand{\shortauthors}{Gonzalez and Matias}

\begin{abstract}
{\color{red} 
\textbf{This document is a pre-print under review as of 10/30/2024 and has not been peer reviewed. Do not share without permission. We do invite you to share feedback on any errors or critical omissions. The best feedback will name a section or line number and include citations.}
\\ \rule{\linewidth}{0.5mm}}

Artificial intelligence and social computing rely on hundreds of thousands of content reviewers to classify high volumes of harmful and forbidden content. Many workers report long-term, potentially irreversible psychological harm. This work is similar to activities that cause psychological harm to other kinds of helping professionals even after small doses of exposure. Yet researchers struggle to measure the mental health of content reviewers well enough to inform diagnoses, evaluate workplace improvements, hold employers accountable, or advance scientific understanding. This systematic review summarizes psychological measures from other professions and relates them to the experiences of content reviewers. After identifying 1,673 potential papers, we reviewed 143 that validate measures in related occupations. We summarize the uses of psychological measurement for content reviewing, differences between clinical and research measures, and 12 measures that are adaptable to content reviewing. We find serious gaps in measurement validity in regions where content review labor is  common. Overall, we argue for reliable measures of content reviewer mental health that match the nature of the work and are culturally-relevant.
\end{abstract}

\maketitle

\section{Introduction}
For over fifty years, operators of digital communication systems from social media to journalism and artificial intelligence have relied on content reviewers to classify information that organizations judge to be so harmful that it must be detected quickly and comprehensively \cite{gillespie_custodians_2018}. The content might include gruesome evidence of war crimes, grotesque confabulations from a generative AI, graphic death threats, or child sex abuse material.
According to decades of worker complaints, reporting by journalists \cite{hardy2011spot, chen2014laborers, mcintyre_behind_2022}, lawsuits\cite{perrigo2022facebook, foxglove2022zuckerberg}, and qualitative research \cite{roberts_behind_2019}, the act of reviewing this content at high volumes under pressure can leave people with substantial and potentially-irreversible psychological harms.

This paper broadly focuses on the psychological experience of ``content review work.'' In this work, a human is repeatedly exposed to potentially harmful material and is expected to classify it without the ability to directly intervene or interact in the situation that generated the material. In theory, reliable measures of content reviewer mental health could inform an understanding of the nature of these harms, their causes, and the effectiveness of interventions to reduce them. Many of the symptoms reported by content reviewers are similar to those of other helping professions such as social work, nursing, and first response. Consequently, measures developed for those fields could potentially be reliably adapted to content review work \cite{spence_psychological_2023}. Such measurements could help technology firms evaluate the working conditions of content-reviewing subcontractors and provide guidance to workers and litigators seeking to enforce compliance with basic labor laws \cite{drootin__2021}. Journalists estimate that the content moderation industry had a value of \$8 billion USD in 2024 \cite{jackson_what_2024}. Companies reportedly pay up to \$500 million per year per contractor for content review work \cite{satariano_silent_2021}, and have paid tens of millions of dollars in settlement awards to content moderators diagnosed with post-traumatic stress disorder \cite{newton_facebook_2020}. Yet only a few very recent scientific studies have set out to measure or improve the well-being of content reviewers, and there is little consensus over what such studies should measure \cite{yang_effects_2023, spence_psychological_2023, spence_content_2024, schopke-gonzalez_why_2022}.

Research and action to support content reviewers is also hindered by the cultural and linguistic range of content reviewing work, a limited supply of mental health professionals, and systemic cultural and linguistic gaps in science. Global technologies require content reviewing expertise from a wide range of regions and languages \cite{york_moderating_2019}, leading to a widely distributed workforce. Furthermore, this labor is traded in a global market that gains efficiencies by allocating it to low-income workers in low-income countries with few labor protections \cite{gray_ghost_2019}. In comparison to countries like the United States, which has 40 mental health professionals per thousand people, many content reviewers live in countries with very limited access to mental health services, including Kenya (0.2 per hundred thousand), India (0.4 per hundred thousand), and Indonesia (0.5 per hundred thousand) \cite{who_health_observatory_2024}. Psychological science itself has similar disparities, with most research focusing on the most well-off groups within wealthy nations \cite{henrich_weirdest_2010}, including research on technology and mental health \cite{ghai_social_2022}. Although surveys of content reviewers report a need for occupational mental health care \cite{spence_content_2024}, it is likely that many of the people who bear the greatest psychological burdens have the least access to occupational mental health services or even the scientific research that could inform those services. 

Efforts to understand and reduce these harms would be substantially aided by reliable measurement of the mental health of content reviewers, especially in the cultures and contexts where people do this work. In this paper, we report the results of a systematic review of psychological measures relevant to content review labor, based on 143 validation studies from fields adjacent to content moderation. We summarize the nature of content review work, outline the characteristics alleged to cause psychological harm, and outline the uses of reliable measurement for designers, organizational leaders, policymakers, and labor advocates. We then include a narrative report on 12 clinical and research measures, a summary of their reported validity, data on the countries and languages where they have been validated, and a review of their potential to inform research on content moderation. We conclude with recommendations for researchers studying content reviewing and mental health.

This paper participates in the tradition of translational systematic reviews in social computing research \cite{stefanidi_literature_2023}. The digital labor of reviewing content is a social psychological human experience created by computing professionals to manage the problems of the computing industry. As such, research on this labor needs to incorporate an understanding of the underlying human-computer interactions while also harmonizing with the state of knowledge in clinical, occupational, and social psychology. This review sets out to bridge those conversations.

\section{Related Work}
In this section, we review prior on content review work to set the context for the use of psychological measures to study reviewer mental health. We then summarize the psychological risks of this work to establish the circumstances where such measures are used. We then summarize the differences between clinical and psychological measures of mental health. Finally, we outline the stakeholders, including designers, corporate managers, and labor unions, who stand to benefit from access to reliable measurement.

\subsection{Content review work}
This paper broadly focuses on the psychological experience of ``content review work,'' in which a human is repeatedly exposed to potentially harmful material and is expected to classify it without the ability to directly intervene or interact in the situation that generated the material. This material could be presented in various media (text, audio, visual, tactile). It could include graphic content, suicide notes, illegal material involving children, misinformation, evidence of war crimes, or other information broadly considered so harmful or risky that it needs to be proactively and urgently classified. We chose the term to describe a specific experience within a wider set of practices that have been called content labeling \cite{morrow2022emerging}, human computation \cite{law_human_2011}, commercial content moderation \cite{roberts_behind_2019}, volunteer moderation \cite{matias_civic_2019}, open source intelligence \cite{fiorella_how_2022}, digital humanitarian response \cite{dubberley_making_2015}, social media reporting \cite{dubberley_finally_2020}, microwork, ghost work \cite{gray_ghost_2019}, and digital crisis response \cite{branson_when_2021}. This general definition offers a bridge to the psychological literature while also highlighting key distinctions between content reviewing and other activities with trauma risks, such as nursing or law enforcement. In this section, we briefly summarize the evolving role of content review work in digital technologies and outline the distinct characteristics of this work that shape the psychological experience of workers.

\subsubsection{How Content Review Work Is Structured}
Scholarship on content review work has outlined multiple layers of relationships and institutions that structure the psychological experience of content reviewers: the user experience, interaction, forms of labor, and response to rapidly changing demand.

At its most basic level, content reviewing work is a \textit{user experience}. A system of bureaucracy and technology compiles materials for review, organizes those materials into a sequence of tasks, and then presents the worker with materials to make judgments on \cite{gray_ghost_2019}. Designers of these systems can control many aspects of the user experience, including the structure of tasks, the rate of tasks, the stimuli presented to workers (blurred or not for example), the rate of exposure, worker agency over task selection, whether tasks are individual or collective, and whether reviewers learn the outcomes of their judgments \cite{steiger_psychological_2021}. The type of content moderators encounter varies in severity, from benign content that was mistakenly flagged to the most extreme material. \cite{steiger_psychological_2021, yang2024perceptions}.

Content review work also varies in the degree to which reviewers \textit{interact} with the situations that they are making judgments about. In systems where content reviewing is one part of a wider set of community leadership and facilitation, people who do review work also have direct contact in real-time with the people and situations they are reviewing \cite{li_all_2022, postigo_emerging_2003, matias_civic_2019}, contact that is associated with differences in well-being \cite{bulat_psychology_2024}. Yet many content reviewers are distanced physically and culturally and separated in time from the situations they review, unable to contact people or intervene in the lives of people who likely do not know the reviewers exist \cite{roberts_behind_2019, gray_ghost_2019}. In situations where content reviewers evaluate the safety and reliability of AI systems, there may be no real situation for intervention at all, and reviewers may be unaware that they are interacting with generated materials \cite{zhang_human_2024}.

The work of content reviewers is structured through widely varying \textit{labor systems} that determine the workload, incentives, penalties, and general working conditions of reviewers. Some workers are employed as trust and safety professionals with salaries, retirement funds, and other corporate benefits \cite{roberts_behind_2019}. Others are volunteers or entry-level staff at journalism and civil society organizations \cite{dubberley_making_2015}. Yet others are prison laborers \cite{lehtiniemi_prisoners_2022, kaun_prison_2020}. While frontline content reviewers are the people exposed to the greatest quantities of extreme content, many other staff are also exposed in their everyday work, including engineers, product designers, and policy teams \cite{tspa2024content}. Many content reviewers work as precarious contractors in high-pressure work environments for companies that are sub-contracted by other firms, often in countries with low minimum wages and few labor protections \cite{buni2016secret, roberts2017social, chen2014laborers}. These working conditions are defined by a competing circle of regulators (who demand content moderation), multiple layers of subcontractors (who sell content reviewing services), and review requesters who are often global corporations seeking to scale digital technologies or mitigate technology harms \cite{gillespie2019custodians}.

Finally, content reviewing offers time-sensitive responses to \textit{rapidly changing demand} shaped by cultural markets, geopolitics, and business competition. Across fields, these dynamics create urgency for large volumes of work at unpredictable times under precarious working conditions: daily quotas ranging from 100 to 1,000 items depending on the platform and type of content \cite{bbc2022moderators,mcintyre_behind_2022,bureau2023dating}. For example, on one platform, moderators must review 900 videos a day with only 15 seconds per video \cite{mcintyre_behind_2022}. While dating app moderators are expected to evaluate a profile every 100 seconds \cite{bureau2023dating}. Moderators for another social media platform are expected to process a ticket every 55 seconds, which leads to between 500 and 1,000 reviews per shift \cite{guardian2024african}. The start or end of a war, natural disaster, or policy change can quickly increase demand for people with no prior experience in content reviewing or lead some reviewers to lose their positions. Similarly, competition between firms for data and AI model validation can create escalating boom and bust cycles for content reviewing labor \cite{ganguli_red_2022}.

\subsection{The psychological risks of content review work}
According to news reports, lawsuits, ethnographic investigations, and some survey research, content reviewers experience similar mental health harms as other helping professionals \cite{spence_content_2024}. Yet some elements of the role, including the pace of material and dissociation from the situation, are less common. In this section, we review the psychological risks to content reviewers through the lens of research on these other professions.

Research with front-line professionals such as first responders has shown that being exposed to other people's traumatic experiences can lead to post-traumatic stress disorder (PTSD) \cite{obuobi-donkor2022,spence_psychological_2023}. This disorder can develop after being exposed to actual or threatened death, serious injury, or sexual violence, either by experiencing or witnessing the event directly or through repeated or extreme exposure to distressing details of the traumatic event(s). It is characterized by intrusion (unwanted, involuntary thoughts), avoidance (avoiding triggering situations or any discussion of the symptoms), negative emotions (such as fear, blame, guilt, shame), and hyper-arousal (irritability, becoming overly watchful of surroundings, difficulty sleeping) \cite{apa2013,spence_psychological_2023}.

People working in helping professions such as social work or therapy can also experience conditions such as secondary traumatic stress (STS), vicarious trauma (VT), and burnout due to their exposure to others' trauma \cite{craig2010,greinacher2019,lee2018,spence_psychological_2023}. These conditions are sometimes called the ``cost of caring'' and are related to the stress experienced in response to another person's distress \cite{figley1995,spence_psychological_2023}. STS shares symptoms with PTSD, such as intrusive thoughts, hyperarousal, and hypervigilance. The harms of VT can grow substantially across a person's lifespan because it changes a person's self-image and view of the world to be less trustful, more fearful, more vulnerable, and cynical \cite{foley2021,krause2009,spence_psychological_2023}. Similarly, burnout is characterized by emotional exhaustion, cynicism, and decreased feelings of accomplishment \cite{brady2017,omalley2019,spence_psychological_2023}.

Researchers have observed similar symptoms across professions that are not traditionally considered front-line staff. For instance, journalists exposed to traumatic events while dealing with distressing content such as child abuse, child cruelty, war, and aviation accidents tend to experience higher levels of PTSD and internalize guilty thoughts \cite{browne2012,spence_psychological_2023}. Professionals who help refugees, asylum seekers, and displaced people also experience adverse mental health effects. These can include Secondary Traumatic Stress (STS) and Vicarious Trauma Stress (VTS) \cite{ebren2022}. A recent study of fifteen research projects showed that forty-five percent of professionals and volunteers working with forcibly displaced people experienced secondary traumatic stress--including intrusive re-experiences of the stories they had heard from refugees \cite{roberts2021,ebren2022}.

Content reviewers, such as caring professionals, journalists, or humanitarian workers, also experience traumatic encounters indirectly. They are often required to view depictions of child sexual abuse, violence, cruelty, humiliation, and discrimination for extended periods \cite{steiger_psychological_2021,spence_psychological_2023}. As a result, they report experiencing anxiety, depression, nightmares, fatigue, and panic attacks, which also impact their relationships and physical health \cite{roberts_behind_2019,newton2019,newton2020b,spence_psychological_2023}. Academic research has highlighted that content reviewers experience high-stress levels, fatigue, insomnia, anxiety, low mood, increased apathy, guilt, and intrusive images as part of their everyday work \cite{dosono2019,lo2018,cook_awe_2022,benjelloun2020,spence_psychological_2023}. Content reviewers may not initially notice this effect, which can be gradual and cumulative, making it challenging for individuals to recognize any changes in themselves or others \cite{krause2009,ledingham2019,spence_psychological_2023}. When content reviewers provided examples of how they were affected, they mentioned experiencing intrusive thoughts and images related to CSAM, often triggered by children and their own sexual behavior \cite{spence_psychological_2023}. Some also talked about avoiding children and experiencing cognitive and emotional effects such as hyper-vigilance around children, anger, increased distrust of others, and symptoms of hyper-arousal, including sleep disturbances and bodily sensations of anxiety \cite{spence_psychological_2023}. 

The psychological consequences of chronic job-related stress or exposure to traumatic content at work can be severe and long-lasting \cite{vasconcelos2021,spence_psychological_2023}. Like people in similar professions, content reviewers have described developing a darker, more cynical view of the world and other people \cite{omalley2019,spence_psychological_2023}. These views may persist even after they leave their job \cite{omalley2019, spence_psychological_2023}. Younger workers tend to experience greater distress after trauma, more symptoms, and higher levels of burnout \cite{acierno2006,adams2006,brady2017,spence_psychological_2023}. Many content reviewers are relatively young, potentially making them more vulnerable to psychological harm while also having more of their lives to bear the consequences.

Researchers have also found that working with hate speech is linked to intrusive thoughts and hyper-vigilance \cite{jereza2022,spence_psychological_2023}. This type of content is the most common content that moderators have to review on social media platforms \cite{facebook2022a,twitter2021,spence_psychological_2023}. Researchers hypothesize that if hate speech is associated with violent or graphic content, moderators' worldviews may be altered, possibly aligning with what they are exposed to \cite{jereza2022,spence_psychological_2023}.

Further research is necessary to determine the prevalence of these harms among content reviewers, whether they vary on the basis of the content that they review, and what can be done to manage these harms \cite{jereza2022,spence_psychological_2023}. To answer these questions, we need larger-scale quantitative and longitudinal studies to measure potential effects and their long-term consequences, quantify potential impacts, and explore possible factors that increase content reviewer resilience \cite{burns2008,samsha2014,spence_psychological_2023}—issues that have been explored qualitatively in CSCW and HCI scholarship, e.g., \cite{dosono2019,finholt_psychology_1998,fox_patchwork_2023,cook_awe_2022}.

\subsection{Clinical and Research Uses for Psychological Measures of Workers}
Clinical and occupational psychologists make important scientific, occupational, and legal distinctions between different kinds of psychological measures. Notably, measures for individual diagnoses (clinical use) differ from those designed to shed light on a work environment or occupation (research use). For clinical psychologists, a reliable measure is a narrow diagnostic tool to inform interventions that could improve their health as well as provide reliable information to employers, insurers, and courts. For research-oriented occupational psychologists, a reliable measure will reveal new understandings about people, their environments, and the nature of their occupations.

Psychological surveys used for clinical purposes are designed to inform diagnoses of conditions that might include mental illness or some other psychological harm, whether or not a workplace experience has caused the damage. A reliable measure will help clinicians decide whether to classify people's experiences with a recognized mental health condition, whether it is short-term and reversible or not. Professional medical and psychiatric organizations, such as the American Psychiatric Association (APA), the World Health Organization (WHO), and the National Institute for Occupational Safety and Health (NIOSH) maintain consensus systems for carrying out these diagnoses, which are recognized by medical professionals, insurers, and courts \cite{who_icd11,apa2013,tetrick2023handbook}. For example, clinical psychologists have added Secondary Trauma as an extension of Post-Traumatic Stress Disorder (PTSD) in the DSM-5, which is used to diagnose mental health \cite{hydon2015preventing,apa2013}.

Many other psychological measures of workers are focused on supporting scientific discoveries about people and their environments rather than supporting clinical diagnoses. For example, the Maslach Burnout Inventory (MBI) was originally developed to understand organizational stress and job-person fit rather than diagnose a medical condition independent of someone's job \cite{MaslachJackson1981,kahn1992stress}. Similarly, measures such as the Copenhagen Burnout Inventory (CBI) seek to measure a person's capacity and ability to manage their mental resources during or outside work periods \cite{kristensen2005copenhagen}. By studying people's psychological experience of work in context, researchers hope to understand the psychology of work better, compare different professions, and offer insights into the effects of different management approaches.

Psychologists insist that research-focused psychological measures should not be used for clinical purposes, which they have not been designed for \cite{tetrick2023handbook}. First, using the wrong measure for the wrong purpose could generate misleading findings if the instrument is measuring something that is not a clinical outcome. Second, non-clinical measures are not designed to support binary decisions about whether someone has a condition or not. So-called diagnoses based on non-clinical measures can be prone to error, especially when people without clinical training make arbitrary decisions about the threshold for diagnosis. Using clinical measures for non-clinical purposes can also cause difficulties for scientists when they lack the statistical properties needed to estimate correlation or causation across groups.

Decisions about what to measure have practical consequences for the people and organizations being measured. Since clinical measures are used to make claims about the health and competence of individuals, people may resist diagnosis due to potential stigma and workplace consequences. This diagnosis avoidance could lead to biased estimates in research studies. Mismeasurement can also lead to misunderstandings of worker conditions. For example, studies that prioritize clinical diagnoses alone may fail to observe workplace conditions that are important factors in a person's experience. Conversely, research-focused measures that incorporate more information about the workplace can sometimes fail to observe important health diagnoses \cite{tetrick2023handbook}.

Because clinical and research measures have different functions and prioritize different things, we differentiate between them in our systematic review. 

\subsection{Applying Psychological Measurements in CSCW}
Psychological measurement has played a central role in the fields of human-computer interaction and computer-supported cooperative work \cite{card_psychology_1986, olson_research_1997, finholt_psychology_1998,norman_design_2013, kiesler_social_1984}. In many cases, designers and organizations can develop bespoke surveys in user experience design \cite{ozok_survey_2007}. Yet high-stakes situations such as mental health rely on scientifically-validated measures.

Researchers in human-computer interaction have called for greater consideration of trauma in the design of technologies, making reference to the work of people who review harmful content for social media platforms \cite{scott_trauma-informed_2023} and artificial intelligence training systems \cite{zhang_human_2024}. Because the design of systems that structure human labor necessarily involves conflicting needs and aims \cite{gregory_scandinavian_2003} with different evidentiary requirements, no single measure of mental health will serve all purposes. In this section, we summarize the different parties who need to measure content reviewers' mental health and their sometimes conflicting use cases for clinical and psychological measures.

Content reviewing labor is structured in a complex supply chain of organizations with conflicting interests around worker mental health \cite{roberts_behind_2019}. This complexity is reflected in the scholarship within social computing, where researchers sometimes operationalize mental health as an independent variable that predicts productivity and retention for unpaid workers \cite{schopke-gonzalez_why_2022}, sometimes describe it as an occupational risk \cite{spence_content_2024}, and other times describe it as a social justice concern \cite{steiger_psychological_2021}. In public talks, trust and safety professionals at large platforms have also described mental health measurement as a tool for supply chain transparency as they evaluate the constellation of contractors and sub-contractors who provide this labor \cite{gilbert_what_2022}. 

\textbf{Content Reviewers}
Many content reviewers may not realize they are experiencing mental health issues, as such problems can manifest gradually \cite{Te-Brake2008} and reviewers may become accustomed to feelings of anxiety, stress, and unhappiness at work and view these emotions as common \cite{Te-Brake2008, capel_longitudinal_1991, spence_psychological_2023}. Ignoring early warning signs can exacerbate negative mental states, leading to illnesses, burnout, and other health problems. Such burnout can also negatively impact job performance, relationships, and home life \cite{shirom_discriminant_2003,spence_psychological_2023}. Self-assessment could help workers prioritize their well-being and observe early signs of discomfort \cite{spence_psychological_2023}. With feedback on their symptoms, workers could make informed assessments of how their work affects them, seek care, and request changes to their work arrangements \cite{steiger_psychological_2021,spence_psychological_2023,roberts__2023}.

\textbf{Organizations that rely on content reviewing}
Since content reviewing work implicates many organizations across a labor supply chain of for-profit, non-profit, and public-interest organizations, these organizations have many potential uses for mental health measurement, some of them conflicting. Within organizations, measurements are relevant to three groups: managers, clients, and organizations that work with volunteer content reviewers.

\textit{Managers and designers} need to be able to identify working conditions that are especially harmful to whole groups of workers, whether those working conditions are the design of a software workflow, the design of an office building, or the management culture of a unit. They also need to be able to evaluate the effectiveness of interventions designed to improve workplace conditions \cite{steiger_psychological_2021}. \text{Clients of organizations} that provide content reviewing labor need to be able to assess the well-being of subcontractors well enough to manage liability and ensure compliance with mental health best practices, sometimes across countries and cultures. Clinical psychologists have reported that some companies have retained them to manage compliance and liability through supply chain transparency, though none of this work has been made public \cite{gilbert_what_2022}. Organizations that work with \textit{volunteer reviewers,} such as humanitarian groups, Wikipedia, or social media platforms, face distinct challenges to sustain levels of volunteering or support people whose trauma may make it difficult for them to step away from content reviewing \cite{mcmillen_wikipedia_2016}. Commercial platforms that rely on volunteer moderation also face this challenge, with the added constraint that too much support for unpaid volunteers might expose them to legal liability if courts determine that moderators should be compensated \cite{postigo_emerging_2003, terranova_free_2012, matias_civic_2019}.

\textbf{Labor Organizers}
Measurements of content reviewer mental health can also be powerful tools for labor organizers. Consider, for example, the publicly stated goals of the Safe Content Advocacy Network (SCAN), led by the Kenyan moderator and organizer Daniel Motaung. This organization seeks to advocate for better working conditions for moderators- an advocacy goal that reliable mental health measures could inform \cite{safe_content_advocacy_network}. These Kenyan workers, alongside content reviewers in Columbia, have now successfully unionized \cite{perrigo_150_2023, mcintyre_teleperformance_2023}. If labor organizers can use mental health measures to show the impacts of employer working conditions, wages, and benefit programs, their advocacy might be placed on a more substantial empirical basis.

\textbf{Social Scientists}
In parallel with the pragmatic interests of tech firms, moderators, and other stakeholders, improved measurements of mental health could also contribute more general discoveries to the fields of psychology and design. Digital communications systems have created a new class of psychological experience for humankind - processing highly disturbing material at an unprecedented scale and speed. Efforts to understand the psychological process created by these design decisions could help us better understand the nature of trauma more generally. Furthermore, attempts to transfer and validate clinical psychology measurements from other fields into content moderation could add clarity to the generalizability and limitations of the standard trauma measures we consider in this review. In parallel, computer scientists have recently developed frameworks for trauma-informed computing design \cite{dellet.altraumainformed}. The lessons we learn about the psychological impacts of moderation could have broader implications for trauma-informed computing, well beyond just content moderation.

\section{Methods}
In this study, we conduct a systematic review of measurements used to assess the mental health of workers and volunteers in fields that are adjacent to content moderation. This systematic review aims to explore current literature on secondary trauma and related phenomena measurement tools and their applicability in the context of content moderation. This paper follows the PRISMA guidelines for systematic reviews \cite{page_prisma_2021}. 

\subsection{Systematic reviews in human-computer interaction}
Systematic reviews play multiple fundamental roles in supporting scholarship in social computing research and ensuring its harmony and impact with the rest of the scientific record. Such reviews have become more common as the body of research in HCI and CSCW has grown and as enduring questions have emerged from decades of the field's existence. Consequently, the number of systematic reviews published in social computing venues has increased tenfold between 2011 and 2021\cite{stefanidi_literature_2023}.

Such reviews are especially common and helpful on questions that overlap with the social, psychological, and behavioral sciences, where ongoing developments in related fields have implications for computer science and vice versa. In recent years, for example, conferences and journals in CSCW and HCI have published systematic reviews on adolescent online safety \cite{pinter_adolescent_2017}, digital well-being \cite{roffarello_achieving_2023}, child well-being \cite{hoiseth_identifying_2017}, the sharing economy \cite{dillahunt_sharing_2017}, and human-robot collaboration \cite{johansen_characterising_2024}. All of these reviews harmonize CSCW and HCI research with the rest of science by summarizing scientific developments outside of computing together with research and perspectives that integrate the contribution of computing to those conversations.



\begin{figure}
    \centering
    \includegraphics[width=0.8\linewidth]{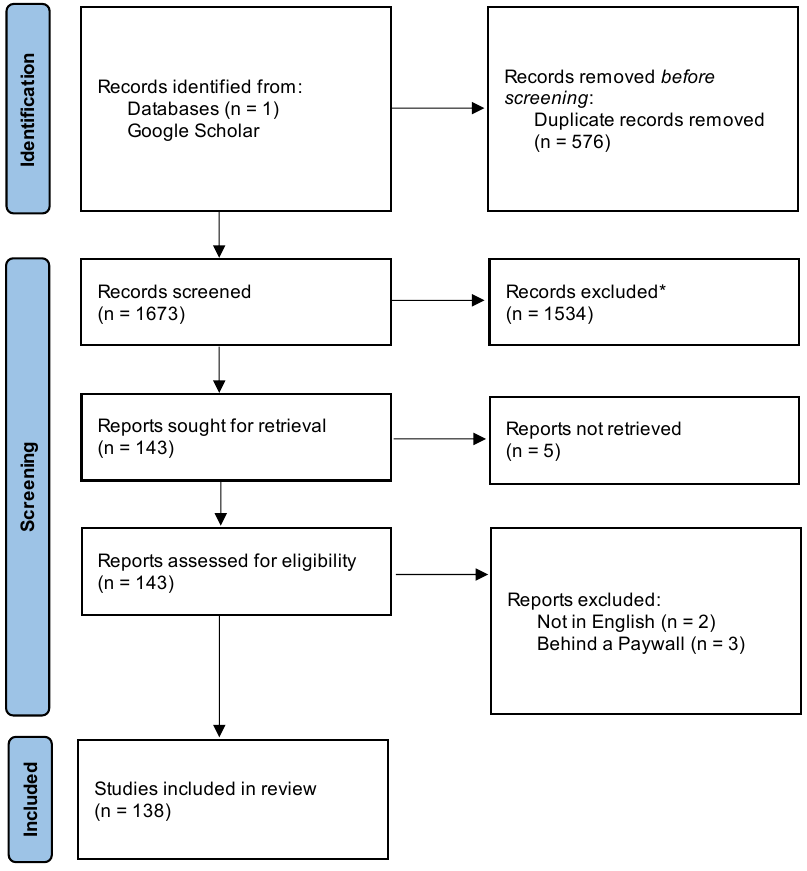}
    \label{fig:PRISMA-DIAGRAM}
    \caption{Selection and filtering criteria for systematic review, following the PRISMA method \cite{page_prisma_2021}.}
\end{figure}

\subsection{Database Identification}
Because research about worker mental health is not limited to medical or psychology journals, we chose to query Google Scholar, which offers a broader range of sources than PsychNet or the National Library of Medicine, including sources across computer science. According to the Google Scholar documentation, it includes references to articles, theses, books, abstracts, and court opinions from academic publishers, professional societies, online repositories, and universities \cite{noauthor_google_nodate}. Importantly for our purposes, Google Scholar also includes whitepapers, technical reports, and other gray literature published by civil society organizations \cite{haddaway_role_2015}, who have also conducted important research on content reviewer mental health. Because Google Scholar sometimes returns different results to the same query, we have included the full results of our search in the supplementary materials.

\subsection{\textit{Selecting Search Terms}}
We searched Google Scholar for key terms linked with mental health, paired with key terms associated with measurement and assessment. Our mental health terms included `burnout,' `secondary trauma,' `vicarious trauma,' `occupational trauma,' and `compassion fatigue.' Our measurement and assessment terms were `measure,' `assess,' and `metric.' Consequently, for each mental health term, we conducted three searches, one for each measurement-related term. For example, we conducted the following searches on burnout: `burnout AND measure,' `burnout AND assess,' and `burnout AND metric.' With four mental health terms and three measurement terms, we conducted a total of twelve searches.

We chose each term for its potential to capture the essence of the phenomenon in question and its usage in various academic and professional contexts. We used the Publish or Perish software to search for the terms and exported the first 150 results from Google Scholar for each combination of search terms  \cite{harzing_-_publish_2016}. We conducted queries on May 29th and 30th, 2024. In total, we collected 2,249 article references. The full list, after removing duplicates, included 1,673 unique articles (Figure 1).

Next, we reviewed the abstracts of each article to evaluate that they were (a) accessible through journals or databases, (b) were published in English, (c) mentioned a quantitative measurement tool, (d) referenced a profession, and (e) attempted to validate the measure and not just apply it.
Out of all of the papers,  143 papers met the inclusion criteria for a full reading, constituting 9\% of papers we considered. We added four papers from the larger dataset to our list of validation studies after noticing later that they also met these criteria.

\subsection{Coding and Grouping Measures by Phenomenon}
A single coder reviewed the final set of 143 papers, recording common information about each paper: the scales tested, whether the paper was a systematic review, whether the measure was available to readers (publicly or upon request), the professional contexts evaluated, what statistics were used to validate the measure, the country where the studied population lived, and the languages used for any survey instruments. We also archived as many survey instruments as were made publicly available. For each of the survey measures mentioned in a paper, we classified them with the psychological phenomenon they were developed to measure, resulting in a set of 184 measures corresponding to  85 psychological phenomena. For example, these included scales measuring job pressure, life satisfaction, anxiety, job turnover, purpose, schizophrenia, mood states, drug abuse, and personality. Many studies in the search mentioned multiple measures since they set out to test for correlations between occupation-focused measures (such as burnout) and other constructs (such as personality).

In this paper, we report on 7 phenomena (Table \ref{table:results}) that we believe are especially relevant to the study of content review labor, either because prior research has studied a phenomenon or because the qualitative and journalistic evidence makes mention of experiences measured by those phenomena. For example, multiple reports have described symptoms of secondary trauma among content reviewers, and at least one study of content review workers has measured it. We have consequently included secondary trauma in our review. In contrast, we have chosen to exclude highly-general measurements such as mood, drug abuse, or work satisfaction. While elements of content review work may correlate with those measures, they do not appear as central priorities in statements from workers, corporate leaders, journalists, or researchers.

\section{Results}
Overall, our search for literature identified 7 psychological phenomena measured by at least one validated scale (Table \ref{table:results}). These relate to a total of 12 measures validated in peer-reviewed articles published between 1981 and 2024. 

In the results, we first group phenomena by their use for clinical or research purposes and report our findings on individual measures alongside others that study similar phenomena. Within each measure we provide a description of the tool and what it measures; report on its validity; and reflect on its potential applicability to content review work.

\begin{table}
\centering
\caption{Selected mental health measurements relevant to the work of content reviewers.}
\centering
\begin{tabular}[t]{llrrrr}
\toprule
Phenomenon & Scale & Year & Papers & Languages & Countries\\
\midrule
\addlinespace[0.3em]
\multicolumn{6}{l}{\textbf{Clinical Measures}}\\
\midrule
\hspace{1em}Secondary Trauma & Secondary Traumatic & 2004 & 12 & 3 & 5\\
 &  Stress Scale (STSS) \\
\cmidrule{1-6}
 & The Traumatic Stress Institute & 1995 & 5 & 1 & 2\\
 & (TSI) Belief Scale  \\
\cmidrule{2-6}
 & Vicarious Trauma Scale (VTS) & 2008 & 4 & 6 & 2\\
\cmidrule{2-6}
\multirow[t]{-3}{*}[2\dimexpr\aboverulesep+\belowrulesep+\cmidrulewidth]{\raggedright\arraybackslash \hspace{1em}Vicarious Trauma} & Trauma and Attachment  & 2003 & 3 & 1 & 3\\
& Belief Scale (TABS) \\
\cmidrule{1-6}
 & Compassion Fatigue-Short Scale & 2006 & 3 & 2 & 2\\
\cmidrule{2-6}
\multirow[t]{-2}{*}[1\dimexpr\aboverulesep+\belowrulesep+\cmidrulewidth]{\raggedright\arraybackslash \hspace{1em}Compassion Fatigue} & Compassion Fatigue Scale & 1995 & 5 & 1 & 3\\
\cmidrule{1-6}
\hspace{1em}Well-Being & WHO Well-Being Index & 1998 & 3 & 2 & 4\\
\cmidrule{1-6}
\addlinespace[0.3em]
\multicolumn{6}{l}{\textbf{Research Measures}}\\
\midrule
 & Maslach Burnout & 1981 & 76 & 23 & 32\\
 & Inventory (MBI) \\
\cmidrule{2-6}
 & Burnout Assessment Tool (BAT) & 2020 & 16 & 14 & 18\\
\cmidrule{2-6}
\multirow[t]{-3}{*}[2\dimexpr\aboverulesep+\belowrulesep+\cmidrulewidth]{\raggedright\arraybackslash \hspace{1em}Burnout} & Copenhagen Burnout  & 2005 & 14 & 8 & 11\\
& Inventory (CBI) \\
\cmidrule{1-6}
\hspace{1em}Vicarious Resilience & The Vicarious Resilience Scale & 2017 & 1 & 1 & 2\\
\cmidrule{1-6}
\hspace{1em}Compassion Satisfaction & Professional Quality  & 1995 & 14 & 5 & 7\\
 (Compassion Fatigue, Burnout) & of Life Scale (ProQoL) \\

\bottomrule
\end{tabular}
\caption{Selected mental health measurements relevant to the work of content reviewers. \textit{Phenomenon} designates the underlying construct that one or more scales set out to measure. \textit{Scale} is the name of a specific scale. \textit{Papers} reports the number of validation studies found in the literature search. \textit{Languages} reports the number of languages that the scale has been validated in, and \textit{Countries} lists the number of countries where a given scale has been validated. This table is based on 143 papers; the sum of the Papers column is higher since some papers validated more than one measure.}
\label{table:results}
\end{table}

\subsection{Clinical Measures}
Clinical measures of mental health are one important tool within the larger process of diagnosis. These measures are typically used in a screening process to determine whether someone should be seen by a professional for further evaluation. They cannot be used alone as proof that someone should receive a given diagnosis. A negative screening result does not necessarily mean that an individual lacks symptoms that warrant intervention \cite{ncbi2023chapter}. Nonetheless, since these measures correlate strongly with diagnoses, they are useful for organizations and researchers seeking to understand mental health risks at individual and aggregate levels. 

In our systematic review, researchers validated measures of four different mental health constructs: secondary trauma, vicarious trauma, compassion fatigue, and well-being. 

\subsubsection{Secondary Trauma} 
The term "secondary traumatic stress" refers to the emotional disruption experienced by individuals who frequently interact with trauma survivors, such as social workers, and may themselves become indirect victims of the trauma \cite{figley1995,bride2004development}. It is also considered a potential job hazard for individuals who directly assist traumatized populations \cite{figley1999compassion,bride2004development,munroe1995preventing,pearlman1999selfcare}. It refers to the emotional and behavioral reactions that occur when someone learns about a traumatic event experienced by another person and wants to help them. Examples of secondary exposure include learning about a close friend or relative's experience or being repeatedly exposed to details of a traumatic event \cite{apa2013,rodenforeman2017secondary}. The adverse effects of secondary exposure to a traumatic event are very similar to those of primary exposure, with the critical difference being that the event is traumatizing for a second person. Symptoms of secondary traumatic stress closely resemble those of post-traumatic stress disorder (PTSD), including intrusion, avoidance, and arousal \cite{figley1999compassion,bride2004development}. In fact, secondary exposure to trauma is now recognized as a valid DSM-5 Criterion A stressor for PTSD \cite{apa2013}.

\textbf{Secondary Traumatic Stress Scale}

The Secondary Traumatic Stress Scale (STSS) \cite{bride2004development} is a 17-item self-report measure of workplace stress that assesses symptoms of secondary traumatic stress and is included in the DSM-5 for trauma-related disorders as of 2013 \cite{apa2013}. Respondents use a five-choice response format to indicate how frequently they experienced intrusion, avoidance, and arousal symptoms over the past week. The STSS consists of three subscales: Intrusion, Avoidance, and Arousal \cite{bride2004development}.

The DSM-5 is used worldwide, and the STSS has been validated in various languages and cultures \cite{apa2023faq}. The validation study of the STSS reported an overall Cronbach's alpha of 0.93, with subscale values of 0.80 for Intrusion, 0.87 for Avoidance, and 0.83 for Arousal \cite{bride2004development, strange2021ask}. Our search included 12 papers that validated the measure since it was first proposed in 2004. Our search included papers that validated this measure in 3 languages across 5 countries.

Evidence from courts, news reports, and qualitative research have provided compelling evidence that commercial content moderators do experience some forms of secondary trauma \cite{roberts_behind_2019}. Yet the STSS measure was designed for clinicians with traumatized clients whose working conditions differ from content reviewers\cite{bride2004development}. The measure includes an assumption that workers can choose cases, which may not hold true for all content reviewers \cite{spence_content_2024,spence_psychological_2023,bride2004development,bride2007}.

\subsubsection{Vicarious Trauma}
Vicarious trauma, which was first observed with trauma therapists \cite{figley1982traumatization,mccann1990b}, has been defined as the emotional residue of exposure to traumatic stories and experiences of others through work. This trauma can result when people witness fear, pain, and terror that others have experienced and become preoccupied with horrific stories told to them \cite{american_counseling_association_vicarious_trauma}. Symptoms can parallel those of PTSD, including re-experiencing, numbness, avoidance, and persistent arousal, all of which can persist for months or years \cite{figley_1996}. Vicarious trauma, which is sometimes referred to as compassion fatigue, secondary traumatization, secondary stress disorder, or insidious trauma, was included in the DSM-5 as part of the cluster of “trauma and stressor-related disorders" as of 2013 \cite{american_counseling_association_vicarious_trauma,apa2013,pearlman_saakvitne_1995}.  

\paragraph{\textnormal{\textbf{The Traumatic Stress Institute (TSI) Belief Scale \& Trauma and Attachment Belief Scale (TABS)}}}\mbox{}\

The Trauma and Attachment Belief Scale (TABS) is a clinical measure of people's beliefs about psychological needs that are particularly sensitive to trauma \cite{varra_factor_2008}. In this model, people's beliefs provide a cognitive schema through which they see the world and interact with others. When vicarious trauma changes these beliefs, their work, personal relationships, and experience of their own self are negatively affected. TABS, an 84-item scale based on the earlier Traumatic Stress Institute Belief Scale (TSI)  \cite{Pearlman1995}, measures five areas of psychological need, including safety, trust in one's self and others, esteem ("the need to feel valued by oneself and others"), intimacy to one's self and others, and control over one's feelings, actions, life, and interpersonal situations \cite{varra_factor_2008}. 

Because TABS and the TSI are built around five areas of psychological need, research that validates the scales has provided overall validity reports alongside results for each of the five areas. Our search included 5 papers that validated the TSI and 3 papers that validated the TABS since they were first proposed in 1995 and 2003. Validation studies of the TSI reported Cronbach's alphas of 0.95 \cite{Jenkins2002} to 0.98 \cite{Pearlman1995}. The overall test-retest reliability for TABS is .75, with subscales ranging from .60 to .79 for test-retest reliability \cite{strange2021ask,pearlman2003trauma}. None of the validation studies in our dataset were conducted in languages other than English.

Almost by definition, content reviewers are repeatedly exposed to high volumes of experiences that are dangerous and traumatic. News stories and academic research have documented how many content moderators, journalists, and other reviewers see the world as a more dangerous and less trustworthy place after starting this work \cite{lockwood2024,catàfiguls2024,spence_psychological_2023}. Conceptually, these measures are applicable to content reviewing with minimal revision. However, our review only found validation studies for TABS and the TSI in English. For these measures to be usable for content reviewing work, they would need further multi-lingual and international validation. 

\paragraph{\textnormal{\textbf{Vicarious Trauma Scale (VTS)}}}\mbox{}\

The Vicarious Trauma Scale (VTS), created by L. P. Vrklevski and J. Franklin, first assessed vicarious trauma in the legal profession \cite{vrklevski_franklin_2008,aparicio_michalopoulos_unick_2013}. The designers of the VTS set out to create a much shorter and more efficient 8-item tool than the 84-item TABS for measuring distress among professionals working with traumatized clients, especially to identify professionals who have higher levels of exposure to distressing material than others \cite{vrklevski_franklin_2008,benuto_singer_cummings_ahrendt_2018}. 

Four of the eight items in this scale involve the nature of a person's work (for example, whether it exposes them to distressing material and experiences). Three of the items involve a person's psychological experience at work (such as ``sometimes I feel helpless to assist my clients in the way I would like.'') One question considers the person's psychological experiences outside of work. All questions are rated on a 7-point Likert scale, from "strongly disagree" to "strongly agree" \cite{vrklevski_franklin_2008}. 

The initial validation study for VTS reported a Cronbach's alpha of 0.88 \cite{vrklevski_franklin_2008, strange2021ask}. Our search included 4 papers that validated the tool since it was first proposed in 2008. We also found in our search that the tool has been validated in 6 languages, and studies were conducted in 2 unique countries.

While the VTS has the advantage of brevity, it is designed with several assumptions about the nature of care work that may not apply to content reviewers. First, the measurement assumes that the surveyed professionals have the opportunity to intervene in meaningful ways, an assumption that is not true for content reviewers. Second, the survey asks questions about differences in levels of exposure to distressing materials and situations, assuming a priori that higher exposure will be associated with greater vicarious trauma. This assumption makes the measure less useful for investigating the relationship between exposure and vicarious trauma or testing the effectiveness of interventions to reduce that trauma. We suggest that researchers attempting to adopt the VTS should re-design and validate it specifically for content reviewing work, as some researchers have recently done \cite{spence_content_2024}.

\subsubsection{Compassion Fatigue}

Compassion fatigue was first used in healthcare in 1992 by a nurse educator, Joinson, to explain the ‘loss of the ability to nurture’ in emergency nurses who experienced tiredness, depression, anger, and detachment due to intense workloads and complex patient needs \cite{joinson1992,SINCLAIR2017}. Early developers of the term and its related measures were seeking a less stigmatized and more `user-friendly' term to describe secondary traumatic stress \cite{bride2007}. Since then, the concept has been adopted by researchers studying other healthcare providers, psychotherapists, and trauma workers \cite{figley1995,SINCLAIR2017}. Since compassion fatigue overlaps with secondary trauma and burnout, scholarly conversations about the differences between them hinge on small, important details, such as whether a scale measures current compassion fatigue or longer-term conditions \cite{bride2007}.
Clinical psychologists agree that measures of compassion fatigue are best used for screening rather than direct diagnosis \cite{stamm2009proqol,SINCLAIR2017,bride2007}.

\paragraph{\textnormal{\textbf{Compassion Fatigue Scale}}}\mbox{}\

Over the past two decades, researchers have created multiple measures of compassion fatigue: the Compassion Fatigue Self Test (CFST) (40 items) \cite{figley1996psychometric}, the Compassion Fatigue Scale-Revised (CFS-R) (30 items broken into secondary trauma and burnout) \cite{gentry2002treating}, and the Compassion Fatigue-Short Scale (13 items also broken into secondary trauma and burnout) which is a shortened and revised version of the CFS-R \cite{adams2006,bride2007}. Some of these scales include elements from other phenomena, including well-being (such as how happy a person is), burnout (how long they plan to do similar work), and secondary traumatic stress (such as avoidance). Overall, compassion fatigue scales prioritize the relationship between a worker in a helping profession and the people they help, with an emphasis on the interactions someone has with people they help, the outcomes of those interactions, and the emotional experience of those interactions. The scale also includes questions about a worker's relationship to their co-workers.

While the earliest validation study for CFS-R's did not report reliability or validity information, follow-up studies found the mean Cronbach's alphas for each sub-scale of secondary trauma and job burnout to be of 0.70 and 0.55 \cite{adams2006compassion,bride2007,gentry2002treating}.
For the CF-Short Scale, the validation study found each sub-scale showed internal reliability, with the Work Burnout scale having a Cronbach's alpha of .90 and the Secondary Trauma scale having an alpha of .80 \cite{bride2007}. Our search included 3 papers that validated the CF-Short Scale since it was first proposed in 2006. We also found in our search that the tool has been validated in 2 language, with studies conducted in 2 unique countries. 

Compassion fatigue measures make numerous assumptions about the nature of helping work that may not apply to content reviewing. Many of the scales assume a two-way interaction between workers and clients, an interaction that does not exist for most content reviewers. Compassion fatigue measures also make assumptions about coworker relationships that might only apply in some cases, especially for people working remotely or from prison. For this reason, we expect that researchers seeking to use compassion fatigue in content reviewing research will need to significantly re-design the measure to account for the unique characteristics of this work.

\subsubsection{Well-Being}

While well-being as a concept has a wide range of non-clinical purposes and measures \cite{VANDERWEELE2020106004}, some measures of well-being have clinical functions to screen for depression \cite{topp2015who5,johansen1998wellbeing}. Consider the WHO-5 Well-Being Index, maintained by the World Health Organization. Although the WHO-5 is not a standalone diagnostic tool, it has been used for screening people for depression \cite{topp2015who5,johansen1998wellbeing}. Consequently, researchers have used well-being as a tool to understand the mental health of workers in helping professions such as nursing \cite{lara2022psychometric}.

Scientists who study well-being make a distinction between measures of someone's psychological states and measures that assess ``any aspect of their life (e.g., finances, physical health)'' \cite{VANDERWEELE2020106004}. While these two categories of well-being measures often correlate with one another and are sometimes used interchangeably, neither category encompasses the other. In this paper, we focus on measures of psychological well-being since researchers who study content review work are often interested in the effect of changes in material conditions on someone's mental health.

\paragraph{\textnormal{\textbf{WHO Well-Being Index}}}\mbox{}\

The WHO Well-Being Index (WHO-5) was introduced in 1998 to assess well-being in primary healthcare patients \cite{johansen1998wellbeing,topp2015who5}. This 5-item scale measures psychological well-being, derived from the WHO-10 and an earlier 28-item study \cite{bech1993rating,bech1996who,warr1985experience,topp2015who5}. Respondents rate 5 positively phrased statements on a 0-5 scale based on the past 14 days, resulting in a score from 0 to 25, which is then multiplied by 4 for a percentage (0-100). A cut-off score of $\leq$	50 is used for screening depression, but the WHO-5 is not a standalone diagnostic tool \cite{topp2015who5,johansen1998wellbeing}.

The WHO-5 is widely used around the world and has been translated into over 30 languages. Because our search criteria focused on occupational harms, we only found 3 papers that validated the tool in those settings. Researchers found that the WHO-5 demonstrated good reliability, with a Cronbach’s alpha ranging from 0.80 to 0.88 \cite{romano2022italian}.  We also found in our search that the tool has validated in 2 languages, and studies were conducted in 4 unique countries. 

In research with content reviewers, scientists have used well-being measurements, including the General Health Questionnaire, the Positive Affect Negative Affect Scale (PANAS), and the Short Warwick-Edinburgh Mental Well-being Scale \cite{spence_content_2024,steiger_psychological_2021,dang_but_2020}. Unlike with the WHO-5, we did not find any studies in our search that mentioned these scales. The WHO-5 also stands out because of its role in clinical psychology and its widespread adoption across cultures \cite{jahoda1958positive,fava1999wellbeing,topp2015who5}. Yet well-being surveys are insufficient on their own for studying content reviewers mental health since well-being and depression are not the only psychological concerns mentioned by workers.

\subsubsection{General Clinical Psychology Measures}

General clinical psychology measures are essential to understand understanding issues like depression, anxiety, and stress. In this review, we focus on more specialized tools such as STSS, VTS, and TABS because we believe they are better suited to capture the specific psychological harms experienced by content moderators. Although content moderators report anxiety, depression, nightmares, fatigue, and panic attacks \cite{newton2019,newton2020b,spence_content_2024}, mental health problems such as PTSD, STS, and burnout also correlate with other diagnoses such as depression and anxiety \cite{spence_psychological_2023,hakanen2008job,schindelallon2010longitudinal}. In some cases, their onsets are linked. One study found that over half of the surveyed content moderators scored high (13 and over) on a general depression scale (CORE-10), indicating clinical depression \cite{spence_content_2024,barkham2013core10}. This rate is higher than in other professions such as policing, where a quarter of the personnel scored in the clinical range for depression due to dealing with child sexual abuse and exploitation material \cite{conway2023protecting,spence_content_2024}. Higher exposure to traumatic content and the continuous nature of digital work may contribute to the higher observed rates of mental health issues among content moderators \cite{barrett2020who,perez2010secondary,doyle2023impacts,coles2014qualitative,roberts_behind_2019}. While general tools are useful for measuring a wider range of adverse mental health effects of content reviewing, this review focuses on measures that could reveal the more unique mental health challenges faced by workers.

\subsection{Research Measures}
Researchers also need to study psychological experiences that may never be classified as diagnosable clinical conditions. In this paper, we also report on research measures designed with goals other than clinical diagnosis \cite{ADPHealth2023,Columbia2023}. For instance, the World Health Organization has stated that burnout is an occupational phenomenon, not a medical condition \cite{WHO2023,LeiterMaslach2024}. Instead, the WHO positions burnout as a condition related to the context in which work occurs rather than an individuals' personal weaknesses or diagnosis \cite{LeiterMaslach2024,WHO2023}. While diagnosing and potentially stigmatizing an individual for a workplace problem would be an ethical and scientific failure, researchers and regulators still need to know about those workplace problems. Similarly, some measurable problems may not be attributable only to the worker or the workplace but an unfortunate convergence of the wrong person in the wrong job \cite{LeiterMaslach2024}. The measurements in this section have all been validated to some degree, even if they are not appropriate for clinical diagnosis.

\subsubsection{Burnout} 
Since occupational psychologists first developed burnout to study professionals in health and human services in 1981, it has become the most-researched measurement of a person's negative experience in their workplace \cite{LeiterMaslach2024,schaufeli2017}. Researchers who developed burnout scales sought to understand why care professionals were disengaging from their work \cite{schaufeli2017}. The WHO defines burnout as a syndrome resulting from chronic workplace stress. It is characterized by feelings of energy depletion, increased mental distance from one’s job, and a sense of reduced professional efficacy \cite{who_icd11,LeiterMaslach2024}. 

Burnout is an occupational phenomenon, not a medical condition, and should not be equated with depression \cite{bianchi2015,LeiterMaslach2024}. It can increase vulnerability to subsequent mental and physical disorders but is not an illness in itself. Consider burnout as a crisis in people's relationships with their workplaces, rather than as a disorder within individuals. Psychologists have observed that burnout isn't proof that a person or their workplace is necessarily flawed. It might be evidence of poor alignment between both parties when expectations, obligations, and demands from both sides are at odds \cite{LeiterMaslach2024}.

Some researchers have used burnout scales to study the experience of content moderators. For example, volunteer content moderators tend to experience burnout \cite{dosono2019}, apathy \cite{lo2018}, under-appreciation, guilt \cite{wohn2019volunteer}, and a lack of safety \cite{blackwell2018harassment,schopke-gonzalez_why_2022}. Just as occupational researchers have used burnout scales to predict reductions in productivity and people's departure from jobs, researchers have done the same with volunteer content moderators \cite{schopke-gonzalez_why_2022,citizensandtechStudyResults}.

\paragraph{\textnormal{\textbf{Maslach Burnout Inventory}}}\mbox{}\

The Maslach Burnout Inventory (MBI), developed in 1981, quantifies burnout through three dimensions: exhaustion (9 items), cynicism (5 items), and efficacy/personal accomplishment (8 items)  \cite{MaslachJackson1981,DeBeer2024}. Items are scored from 0 "never" to 6 "every day." Initially developed for human services, adapted versions include the MBI-Human Services Survey, MBI-Educators Survey, and the MBI-General Survey (16 items), which is recommended for general workplaces \cite{MaslachJackson1986,SchaufeliLeiterMaslachJackson1996,DeBeer2024}. The MBI-GS has become the primary tool for measuring burnout\cite{DeBeer2024}.

A recent meta-analysis published after our search reported on the validity of the MBI-GS, finding that a third of studies validating the MBI achieved a rating of ``sufficient'' internal consistency using criteria from Consensus-Based Standards for the Selection of Health Status Measurement Instruments (COSMA) \cite{DeBeer2024}. This finding has put the reliability of the MBI into question. Similarly, they found that while the MBI-GS has been translated into a range of languages, it has not been thoroughly validated for comparison between workers within and from different cultures. Overall, the authors report that ``the structural validity of the MBI-GS remains unclear, and so does its cross-cultural validity'' \cite{DeBeer2024}. Our search included 76 validation studies in 23 languages, and studies were conducted in 32 unique countries.

\paragraph{\textnormal{\textbf{Burnout Assessment Tool}}}\mbox{}\

In 2020, Schaufeli et al. introduced the Burnout Assessment Tool (BAT) to improve upon the Maslach Burnout Inventory (MBI) \cite{schaufeli2019bat}. The BAT consists of two sub-scales: BAT-C, focusing on cognitive aspects, and BAT-S, addressing somatic/physical symptoms. Unlike the MBI, which separately examines burnout components (such as cynicism and exhaustion), the BAT views them as connected \cite{schaufeli2023burnout}. BAT-C measures exhaustion dimensions, mental distance, and cognitive and emotional impairments, while BAT-S assesses psychological issues such as tension and psychosomatic symptoms, including headaches.\cite{schaufeli2020bat}. With its 23-item, 5-point Likert scale format, the BAT provides a more comprehensive view of burnout than the 16-item MBI-GS \cite{schaufeli2019bat}.

In the paper introducing the BAT in 2020, researchers reported a Cronbach’s alpha for the BAT-C of 0.95, with subscale alphas for exhaustion at 0.92, mental distance at 0.91, cognitive impairment at 0.92, and emotional impairment at 0.90. The composite BAT-S has an alpha of 0.90, with the alphas for psychological complaints at 0.81 and psychosomatic complaints at 0.85  \cite{schaufeli2020bat}. Scientists have invested significant effort in validating the BAT since it was developed four years ago. Our search included 16 papers that attempted to validate the tool across 14 languages and 18 unique countries. 

\paragraph{\textnormal{\textbf{Copenhagen Burnout Inventory}}}\mbox{}\

The Copenhagen Burnout Inventory (CBI), created in 2005, measures burnout across three sub-dimensions: personal, work-related, and client-related burnout. The 19-item questionnaire focuses on fatigue and exhaustion. Personal burnout questions apply to everyone, while work-related questions assume the respondent is in paid employment. Client burnout questions target those working directly with clients. The development of the CBI aimed to reduce American influence in its wording to better address the needs of a diverse global workforce \cite{CBI2005}.

While our search did not find any studies that meta-analyzed more than one validation study of the CBI \cite{shoman_psychometric_2021}, the initial validation study for the CBI reported Cronbach's alphas ranging from 0.85 to 0.87 \cite{CBI2005}.
Our search included 14 papers that validated the tool since it was first proposed in 2005. We also found in our search that the tool has been validated in 8 languages, and studies were conducted in 11 unique countries. 

\paragraph{\textnormal{\textbf{The Use of Burnout as a Construct in Research about Content Reviewers}}}\mbox{}\

Burnout is a widely-used concept that seeks to measure and explain people's negative experiences in the workplace. As a result, it has also been the most widely used measure to understand the experiences of both commercial and volunteer content reviewers. Yet many of the items in the MBI and related scales are designed for helping professions, asking questions like ``I have the impression that my patients/clients make me responsible for some of their problems'' that are not relevant to all kinds of content reviewing work. Furthermore, many attempts to validate the MBI-GS have produced results that were below commonly accepted criteria for validity, leaving some questions about its reliability and generalizability \cite{DeBeer2024}. Yet science and society would benefit from reliable measurement of the factors in someone's workplace that correlate with the fatigue, exhaustion, and cynicism that helping professionals experience. Future non-clinical, non-policy research about content review work could benefit from further adaption and validation of burnout measures, especially those designed to work across the greatest number of cultures and languages.

\subsubsection{Compassion Satisfaction}

Not all psychological outcomes of care work are negative. Compassion satisfaction is the pleasure derived from helping others, and it has been found to correlate positively with resilience, which is the ability to cope, learn, and grow from difficult experiences \cite{burnett2015compassion,stamm2009proqol,dehlin_lundh_2018}. Compassion satisfaction and compassion fatigue (CF) do not correlate negatively and can coexist \cite{barr_2017,dehlin_lundh_2018}. For example, a psychologist or social worker may experience compassion fatigue and compassion satisfaction simultaneously, but as compassion fatigue increases, it may overwhelm their ability to experience compassion satisfaction \cite{bride2007,dehlin_lundh_2018}. In high-risk work environments, some clinicians experience high levels of compassion satisfaction and secondary traumatic stress (STS) while reporting low levels of burnout. This combination is often seen in highly effective clinicians who feel their work is meaningful but experience significant psychological costs to that work \cite{stamm2009proqol,dehlin_lundh_2018}. Consequently, compassion satisfaction may not always be healthy if it motivates workers to take on psychological harm for work they find satisfying. 

\paragraph{\textnormal{\textbf{The Professional Quality of Life Scale (ProQoL)}}}\mbox{}\

Compassion satisfaction is one part of the larger Compassion Fatigue Self-Test, which evolved into the Professional Quality of Life scale (ProQoL-5)\cite{figley1999compassion, stamm2009proqol}. ProQoL includes subscales for compassion satisfaction alongside burnout and secondary traumatic stress. three 10-item sections \cite{stamm2009proqol}. Respondents rate symptoms on a 5-point scale over the past 30 days, reflecting both positive and negative aspects of their work experience \cite{stamm2009proqol,Geoffrion2019}. The compassion satisfaction components of the ProQoL ask about participants' feelings about the people they are able to help, how proud they feel about their work, and their satisfaction from being able to help others, among others.

In the initial validation study, Cronbach’s alpha for the Compassion Satisfaction component was 0.87 \cite{stamm2005proqol,strange2021ask}. Our search included 14 papers that validated the tool since it was first proposed in 1995. We also found in our search that the tool has been validated in 5 languages, and studies were conducted in 7 unique countries. 

Qualitative research and journalism on content review work, especially volunteer content moderation and humanitarian workers, has reported that people find meaning and purpose in that work when they feel like it is helping others. People also report that this sense of satisfaction can drive them to unhealthy and harmful levels of work \cite{menking_heart_2015, mcmillen_wikipedia_2016, dubberley_making_2015}. Yet existing measures of compassion satisfaction are oriented towards professions where people have direct interactions with clients rather than content reviewing, which is often distanced from the people involved. One research team has used the ProQoL to survey volunteer content moderators for Facebook Groups and subreddits to study why they quit. Yet the findings omit compassion satisfaction results for unknown reasons \cite{schopke-gonzalez_why_2022}. Overall, compassion satisfaction measures could be very informative for future research about the psychological outcomes of content review work and the causes of psychological harm if researchers can adapt the measures reliably.

\subsubsection{Vicarious Resilience}

Many measures of the psychological effects of care work are built on the assumption of a harmful dose-response between work experiences and mental health. Yet some people thrive in high-stress work that supports others. Vicarious resilience is a concept developed by mental health professionals to study the possibility that working with trauma survivors had the potential to positively affect and transform therapists, especially those working with human rights activists, displaced populations, and survivors of torture \cite{gangsei2004torture, hernandezwolfe2018vicarious}. In their qualitative and statistical research, participants referenced the inspiration and strength they drew from working with clients, whom they sometimes described as "heroes" \cite{gangsei2004torture,hernandezwolfe2018vicarious}.

Scholars of vicarious resilience argue that it emerges in the dialog between clients and therapists in ways that promote personal growth among therapists. In this model, therapist and client relationships are interactive, with both parties influencing each other and creating meaning \cite{anderson2007postmodern,hernandezwolfe2018vicarious}. By focusing on the reciprocal nature of the therapeutic relationship, scientists study how therapists themselves can learn from and change alongside their clients \cite{hernandezwolfe2018vicarious}. 

\paragraph{\textnormal{\textbf{The Vicarious Resilience Scale}}}\mbox{}\

The Vicarious Resilience Scale (VRS), developed in 2017, measures the positive impact of clients' recovery on therapists' resilience across seven dimensions \cite{hernandez2007vicarious,hernandezwolfe2014vicarious,engstrom2008vicarious,edelkott2016vicarious,Killian2018}. Its indicators include changes in life goals, client-inspired hopes, self-awareness, self-care, resourcefulness, recognition of clients' spirituality, awareness of power dynamics, and attentiveness to clients' narratives \cite{hernandez2007vicarious,hernandezwolfe2018vicarious,Killian2018}. Originally consisting of 48 items, the VRS was refined to 27 items across a 6-point response scale \cite{Killian2018,sova2021solution}.

The Vicarious Resilience Scale (VRS) has been validated in a single study in two languages, with an overall Cronbach's alpha of .92. The seven subscales had the following alpha scores: increased resourcefulness (.86), changes in life goals and perspectives (.88), increased self-awareness and self-care (.83), client-inspired hope (.80), recognition of spirituality as a client source (.79), increased consciousness (.84), and capacity to stay present during trauma narratives (.65) \cite{Killian2018}.

It is possible that some content reviewers are more resilient than others to the psychological harms of their work, and that some people even grow in resilience over time. However, since most content reviewers are cut off from the situations they classify, questions in the VRS do not match the experiences of content reviewers. The VRS may nonetheless be a useful tool to measure resilience among content reviewers who do have a two-way dialog with the people and situations they review.

\section{Discussion}
In this systematic review, we set out to identify psychological measures used to study the mental health of workers similar to content reviewers, assess the relevance of those measures to content reviewing, and make recommendations for scientific progress on reliable measurement. Within the total set of 184 measures, we report on seven clinical measures and five research measures from related occupational fields.

Clinical measures of mental health play a central role in research and policy. Screening results or diagnosis from a certified clinician can influence a person's access to healthcare, their access to alternative work arrangements, and their potential access to redress in courts if their health needs are not addressed. Clinical measures are also important for companies and managers, since they provide the clearest indicator that someone needs care. 

In our review, clinical measures are available for some harmful psychological experiences that moderators report. Measures of depression such as the WHO Well-Being index are widely translated and validated but do not record information about the kind of intrusive thoughts, avoidance of triggering material, and hyper-arousal that is associated with post-traumatic stress. Although secondary trauma and vicarious trauma are included in the DSM-5 manual for clinical diagnosis, the screening measures for these conditions are not fully-suited to content reviewing work. Measures including the TSI, VTS, TABS, and STSS are designed around low-volume client interactions rather than high-volume work that is often divorced from the people and situations being reviewed. If these measures are applied uncritically to content review labor, they might mis-diagnose or improperly estimate the rates of secondary and vicarious trauma. Similarly, we conclude that compassion fatigue as currently measured is best limited to volunteer and/or community moderation, where workers are allowed to choose which cases to take up, how much to work, and how much to interact with the people in question.

Our review of research measures found multiple scales designed to study the relationship between workers and their working conditions. While these measures cannot provide screening for clinical conditions, they can help designers, managers, and organizers understand a broader range of factors that might correlate with departures from volunteer and paid content review work. Since measures of burnout vary widely in their internal validity, we encourage research teams to review the literature on potential measures before selecting a given scale. Measures of burnout, vicarious resilience, and compassion satisfaction all face a similar validity risk as secondary/vicarious trauma: they were not designed to study high-volume exposure to disturbing material, detached from any chance to help with the people and situations they address. As a result, we encourage strong care when measuring these constructs to adapt and validate versions that are closer to the work of content reviewers.

Beyond these clinical and research scales, a growing body of evidence points to psychological experiences and workplace experiences that are not well measured by current scales because the psychological experience and work conditions are outside what scale-developers initially considered. For example, while many surveys of secondary trauma and burnout have been largely designed to study study professionals with formal training in their profession, content reviewing is often a low-wage, entry level job under precarious labor conditions that is taken up as temporary work. Furthermore, many content reviewers do their work from home at all hours \cite{mcintyre_behind_2022,mcintyre_teleperformance_2023, time2023tiktok}. Despite being surrounded sometimes by family, neighbors or friends, these remote positions can further isolate people from peers or colleagues due to the sensitive nature of their work. Future efforts to improve measurement scales need to account for these aspects of content reviewing.

With the exception of the WHO Well-Being Index, none of the scales have been tested and validated in a range of countries and languages as wide as the regions where people do content review work. We strongly discourage researchers and organizations from conducting comparative analyses between countries and cultures using scales that have not been validated for that purpose. For example, it would be invalid to make comparisons between online platforms or content moderation vendors in different countries on the basis of burnout scores that are not validated for such comparisons. 

\section{Conclusion}
There is no doubt that for many people and workplaces, content review work has caused serious, harmful mental health effects for half a century. Evidence from qualitative research, news reports, lawsuits, and a growing body of scientific literature all describe severe mental health outcomes that companies are also seeking to manage and reduce \cite{gilbert_what_2022}. As society seeks to understand and reduce the human toll of work that is increasingly in demand, everyone will benefit from studies that develop and adopt reliable measures of that toll.

Due to the evidence we have compiled of long-term ignorance from technology firms and scientists alike, we conclude that a genuinely good-faith effort to study and improve the mental health of content reviewers cannot be a small endeavor. For clinical interventions, this lack of measurement is matched with an extreme unavailability of mental health care. Even a modest endeavor to support the mental health of content reviewers will require a sustained, multi-year effort at improving research quality, expanding access to mental health services, and adapting workplace practices in regions that have been most ignored by science and the wealthy world.

These improvements are urgently needed as the demand for content review work grows and as traditional caring occupations are converted into platform labor. In 2024, content reviewing was estimated as an \$8 billion USD industry, with exponential growth expected due to increasing regulation of internet child safety and growth in generative AI \cite{jackson_what_2024}. Furthermore, as mental health care itself has become ``platformized'' through suicide hotlines and teletherapy, a growing percentage of workers in traditional caring and helping professions will find themselves in front of screens handling large quotas of people who they cannot interact with \cite{garofalo_doing_2024}. As that happens, long-standing clinical and research scales will become less and less reliable as screening tools and measurements.

As a matter of primary urgency, we encourage researchers doing descriptive research on content moderation and mental health to prioritize clinical measures and partner with clinical psychologists who are qualified to combine measurement with diagnosis. Up to now, research on the psychology of content moderators has used a range of clinical and research methods with varying levels of validity. Some studies have used measures that include secondary and vicarious trauma \cite{spence_content_2024}. Others have used research-oriented measures of burnout \cite{schopke-gonzalez_why_2022} or some of both \cite{yang_effects_2023}. We hope this review provides researchers with initial guidance on where to look for measures that most closely match their research goals.

Researchers focused on developing interventions to reduce psychological harms to content reviewers should support the development of scales that provide a valid indicator that someone's life has genuinely improved. Across our systematic review, scientists focused on measuring occupational health often wrestled with the challenge of creating measures that were consistent within subjects over time while also responding accurately to improvements. For example scientists cited the difficulty of choosing time-frames in measures of compassion fatigue \cite{bride2007}. If researchers asked people about the last 14 days, they could create measures that responded quickly to short-term interventions. Yet these short term measures sometimes over-estimated changes in experiences that unfold over longer time-scales. We encourage designers, organizational leaders, and intervention scientists to collaborate with clinical psychologists when selecting outcome measures that can reliably detect meaningful changes in content reviewers' mental health.

Because the use of clinical measures can rightly create competing risks for technology operators, researchers, and affected communities, we expect that measurement and improvement of content reviewer mental health will require collective leadership. This includes a need for creative approaches to research that enable scientists to access workers, maintain an ethical duty of care, and protect privacy while managing relationships across technology firms, labor unions, and other stakeholders. Fortunately, other domains of occupational health have developed workable arrangement that balance the needs and incentives of parties that are not always in agreement \cite{kalman_ethical_1999, butler_employer_2017}. We encourage researchers to approach research about content reviewers with an awareness of these best practices.

Finally, we urge all involved parties to use science to accelerate the provision of reliable improvements in the work and mental health of content reviewers rather than slow it down. In other areas involving technology and mental health, scientists' commitment to rigorous evidence has engaged with a pernicious cycle of hype and panic that has failed to deliver measurable public goods over decades \cite{orben_sisyphean_2020}. In particular, the difficulties of measuring content reviewer mental health should not be a reason to delay the design and evaluation of interventions. In other domains seeking to rapidly understand and address severe harms, researchers have used cycles of qualitative methods, descriptive measurement, and field experiments to more rapidly develop actionable knowledge to improve people's circumstances \cite{levy_paluck_promising_2010}. We encourage scientists and institutions to try similarly-accelerated approaches to safeguarding the mental health of content reviewers.

\bibliographystyle{ACM-Reference-Format}
\bibliography{bibliography}

\end{document}